\documentstyle[preprint,tighten,prd,aps]{revtex}
\begin{document}
\preprint{OCIP/C-93-1; TIFR-TH/93-09}
\draft

\title{{\bf EXTRACTING W BOSON COUPLINGS FROM THE\\
$e^{+}e^{-}$ PRODUCTION OF FOUR LEPTONS}}
\author{PAT KALYNIAK, PAUL MADSEN, and NITA SINHA}
\address{{\em Ottawa-Carleton Institute for Physics\\
Physics Department, Carleton University\\
1125 Colonel By Drive\\
Ottawa, CANADA K1S 5B6}}
\author{RAHUL SINHA}
\address{{\em Theoretical Physics Group\\
Tata Institute of Fundamental Research\\
Homi Bhabha Road, Bombay 400 005, INDIA}}
\date{\today}

\maketitle

\begin{abstract}
We consider the processes $e^{+}e^{-}\rightarrow
\ell^{+} \ell^{\prime -} \nu
\bar{\nu}^{\prime}$, including all the possible charged lepton combinations,
with regard to measuring parameters characterizing the $W$ boson. These
reactions
all proceed via virtual $W$ pair production as well as a number of
undistinguished s- and t-channel modes. In addition, some of the processes
also have contributions from other diagrams of interest, those which contain
the $\gamma W W$ or $Z W W$ vertices with gauge bosons in the
t-channel.
  Consequently, the
processes are sensitive to anomalous couplings such as $\kappa_{\gamma}$ and
$\kappa_{Z}$. We here calculate at what level these processes can be used to
measure these anomalous couplings for the cases of $e^{+}e^{-}$ colliders at
500
$GeV$ and 1 $TeV$ center of mass energies. Further, we present helicity
information which should be useful in distinguishing between deviations of
$\kappa_{\gamma}$ from its standard model value and deviations of $\kappa_{Z}$.
\end{abstract}

\section{Introduction}

The gauge boson couplings of the standard model of electroweak
interactions are only just beginning to be directly measured.
There has now been observation of the process $p\bar{p} \rightarrow e
\nu \gamma X$, presumably representing $W \gamma$ production and radiative $W$
decay, at the Collider Detector at Fermilab (CDF) \cite{1} and at UA2 at CERN
.\cite{2} In principle, indirect evidence regarding the gauge boson couplings
comes from higher order corrections to low energy measurements. However, it
appears that the sensitivity to such loop-induced effects of the trilinear
gauge
boson interaction has been overestimated in much of the literature. \cite{3,4}
There now exists some preliminary work on a global analysis of low energy data
and LEP data, in order to extract bounds on the gauge boson couplings;
\cite{3,5}
the  present results are model dependent and incomplete and should be refined.

 The prospect of increasing accumulated luminosity at existing
facilities and of future facilities encourages detailed work on the means of
constraining the gauge boson couplings. We focus here on the possibility of
measuring parameters relevant to the $\gamma W W$ and $ Z W W$ vertices. The
couplings of W bosons to the photon and Z can be described in general by an
effective Lagrangian with seven parameters for each of the neutral gauge
bosons.
\cite{6,7} We will here neglect CP violating parameters as they are constrained
to be less than $ O(10^{-4})$ by neutron electric dipole moment measurements
\cite{8}. An effective Lagrangian respecting CP, C, and P invariance is often
parametrized as  \begin{equation} L=-ig_{V}(g^{V}_{1}(W_{\mu \nu}^{\dagger}
W^{\mu} - W^{\dagger \mu} W_{\mu \nu})V^{\nu}+\kappa_{V}
W^{\dagger}_{\mu}W_{\nu}V^{\mu \nu}
+\frac{\lambda_{V}}{M_{W}^{2}}W^{\dagger}_{\lambda \mu} W^{\mu}_{\nu} V^{\nu
\lambda}) \end{equation} In the above equation, $V$ represents either the
photon
or the Z boson and the overall couplings are taken as $g_{\gamma}=e$ and
$g_{Z}=e \cot\theta_{W}$. The parameters $\kappa_{\gamma}$ and
$\lambda_{\gamma}$ are related to the static magnetic dipole and electric
quadrupole moments $\mu_{W}$ and $Q_{W}$, respectively, of the W boson as
follows. \begin{eqnarray}
\mu_{W}&=&\frac{e}{2M_{W}}(1+\kappa_{\gamma}+\lambda_{\gamma})\nonumber\\
Q_{W}&=&-\frac{e}{2M_{W}}(\kappa_{\gamma}-\lambda_{\gamma}) \nonumber
\end{eqnarray} The tree level standard model values of the parameters of
equation (1) are $g^{V}_{1}=1$, $\kappa_{V} = 1$, and $\lambda_{V} = 0$. If the
W bosons are composite objects, then deviation of the triple gauge boson
coupling parameters from their standard model values could be very large
indeed;
as an example, $\kappa$ has been calculated to be greater than three in one
model. \cite{9} However, within the standard model, upper bounds on the one
loop
corrections to the tree level values of $\kappa_{\gamma}$ and
$\lambda_{\gamma}$
have been given as follows \cite{10} \begin{eqnarray} (\Delta
\kappa_{\gamma})_{max} & = & 1.5\%\nonumber\\ (\Delta \lambda_{\gamma})_{max} &
= & 0.25\%\nonumber \end{eqnarray} In extensions of the standard model such as
those containing extra Higgs doublets, extra heavy fermions \cite{10}, or SUSY
extensions \cite{11} the deviations from the tree level standard model values
tend to be of about the same order of magnitude as these one loop corrections.
Also, $\Delta \lambda$ is usually (although not always) smaller than $\Delta
\kappa$ by close to an order of magnitude, bringing it below a per cent. Hence
we will here neglect deviations of $\lambda$ from its standard model value of
zero and will present numerical results where $\kappa_{\gamma}$ and
$\kappa_{Z}$
vary only within 10\% of 1.

We investigate a set of processes of four lepton production in
$e^{+}e^{-}$ collisions with respect to their sensitivity to gauge boson
coupling parameters. The processes are all of the general form
\begin{equation}
e^{+}e^{-}\rightarrow \ell^{+} \ell^{\prime -} \nu \bar{\nu}^{\prime}.
\end{equation}
Our work
includes all possible charged lepton combinations, specifically these are $\mu
\tau$, $\mu e$ ($\tau e$), $\mu \mu$ ($\tau \tau$), and $ee$. The channels
given
in brackets have the same set of Feynman diagram contributions as their
corresponding unbracketed channel and we will henceforth drop reference to
them as distinct processes.

In the next Section, we describe the four types of processes with respect to
their dependence on $\kappa_{V}$. We discuss our calculations in Section 3 and
present results for the case of unpolarized beams. In Section 4, we present
helicity amplitude information which is relevant to distinguishing
$\kappa_{\gamma}$ and $\kappa_{Z}$ effects. Finally, we summarize our results.

\section{The Four Lepton Processes}

The reactions (2) can all proceed via real
or virtual $W$ pair production, with the subsequent $W$ decays into the
appropriate leptonic modes. The form of the $W$ pair production diagram
which is of interest to our study of the triple gauge boson vertices is
illustrated in Fig.1. For the $\mu \tau$ final state, the diagrams of the
type in Fig. 1 with the $\gamma W W$ and $Z W W$ vertices are the only
interesting ones although they are accompanied by seven additional diagrams.
For
all our processes, we do include the full gauge invariant set of diagrams. The
$\mu e$ final state receives contributions from a total of eighteen diagrams,
including the two W pair diagrams along with fourteen additional rather
uninteresting diagrams. The remaining two Feynman diagrams which contribute are
of interest in the study of the trilinear gauge vertices. They contain the
$\gamma W W$ and $Z W W$ vertices, respectively, with a $\gamma$($Z$) and a $W$
in the t-channel; their form is shown in Fig. 2a. Consequently, by fully
calculating the $\mu^{+} e^{-} \bar{\nu}_{e} \,  \nu_{\mu}$ production, as
opposed to W pair production only, we aim to unearth a more realistic picture
of the sensitivity to the couplings in question. Similarly, for the
process $e^{+} e^{-} \rightarrow \mu^{+} \mu^{-} \nu \bar{\nu}$, there are two
diagrams containing the $\gamma W W$ and $Z W W$ vertices, respectively, in
addition to the W pair diagrams. The form of these contributions is shown in
Fig. 3 and has the $W$ bosons in the t-channel coupling to a photon or $Z$
which decays leptonically. The $\mu \mu$ process has a total of 28 contributing
diagrams with most of the extras being $\gamma$ or $Z$ `bremsstrahlung' from
the
initial or final state leptons. Finally, the process $e^{+} e^{-} \rightarrow
e^{+} e^{-} \nu \bar{\nu}$ goes via a total of 56 diagrams. All the diagrams
containing the triple gauge boson vertices in Figs. 1, 2a, and 3 contribute as
does the diagram of Fig. 2b. For the $\mu \mu$ and $e e$ final states, in some
of the diagrams, all $\nu$ species can appear. These diagrams are added
incoherently in the calculation. However, for the purpose of counting the
number of diagrams, we regard all the $\nu$ final states as contributing to a
single diagram.

While discussing the set of four types of processes, we will also note here
their helicity characteristics. In the following, we will denote the helicities
of the particle set $e^{+}e^{-}\ell^{+}\ell^{\prime -}$ as $(\bar{\alpha}
\alpha \bar{\beta} \beta)$. Fig. 1 contributes to all the processes and goes
via the helicity amplitudes $(+-+-)$ and $(-++-)$ so each process we consider
has these amplitudes. The $(+-+-)$ helicity is actually dominant in all cases.
For the $\mu^{+} \tau^{-}$ final state, no other helicity amplitudes are
introduced among the remaining seven diagrams. Fig. 2a has contributions
from $(+-+-)$ and $(++++)$ helicity amplitudes; thus, the $\mu^{+} e^{-}$ final
state has three helicity amplitudes contributing. For the $\mu^{+} \mu^{-}$
final state, the diagram of Fig. 3 contributes helicities $(+-+-)$ and
$(+--+)$; in addition, some of the extra diagrams without the $\gamma W W$ or
$Z
W W$ vertices have a $(-+-+)$ amplitude. Thus the $\mu^{+} \mu^{-}$ process has
four helicity amplitudes; the $(-+-+)$ amplitude is independent of
$\kappa_{\gamma}$ and $\kappa_{Z}$. Fig. 2b contributes $(+-+-)$ and
$(----)$ amplitudes to the $e^{+}e^{-}$ process. The $e^{+}e^{-}$ process
actually goes via all six possible helicity amplitudes; again, as in the
$\mu^{+} \mu^{-}$ case, the $(-+-+)$ amplitude is $\kappa_{V}$ independent,
arising only in diagrams which do not contain the triple gauge boson vertices.

\section{The Calculations for Unpolarized Beams}

In order to deal easily with the large number of Feynman diagrams and to
readily retain helicity information, we have written the amplitude for each
process in the CALKUL helicity formulation. \cite{12} We assume massless
spinors
describe the fermions although we do retain fermion masses in the propagators;
this amounts to neglecting terms proportional to $m_{f}$, a good approximation.
The matrix element squared for each process is embedded in a Monte Carlo
algorithm for integration over the  final state four body phase space to yield
the cross sections and various distributions. We sum and average over initial
spins and sum over final spins. We use
$M_{Z} = 91.196 \, GeV$, $\Gamma_{Z} = 2.534 \, GeV$, $M_{W} = 80.6 \, GeV$,
$\Gamma_{W} = 2.25 \, GeV$, $m_{e}= 0.511 \, MeV$, $m_{\mu} = 0.1057 \, GeV$,
$m_{\tau} = 1.7841 \, GeV$, and $\sin^{2} \theta_{W} = 0.23$.

We have performed a number of checks on our calculations. We have checked our
algorithms by showing that our $\mu \tau$ results reduce to those of $W$ pair
production \cite{6,13} if only the appropriate contributions are included; this
included checking that the individual contributions from the three W pair
diagrams, $M_{\gamma\gamma}$, $M_{ZZ}$, and $M_{\nu \nu}$ and those from their
interferences $M_{\gamma Z}$, $M_{\gamma \nu}$, and $M_{\nu Z}$ were
reproduced properly. Another useful check on our matrix elements is that of
charge conjugation; we generated various redundant distributions for the
positively and negatively charged leptons for the $\mu^{+} \mu^{-}$, $e^{+}
e^{-}$, and $\mu^{+} \tau^{-}$ (invariant up to the $\mu$ $\tau$ mass
difference) channels as a check. In addition, we generated a number of
distributions which are not actually experimentally observable for our
processes due to the two neutrinos, such as the angular and invariant mass
distributions of the reconstructed W bosons in order to note their consistency
with W pair production work. \cite{6,13}

The experimental signature for the processes under consideration is a clean
one,
an oppositely charged lepton pair and missing transverse momentum and energy
due to the neutrinos. We have made some fairly simple cuts as described below
to account for detector acceptance and potential backgrounds. For all the
processes, we require a cut on the angle of each of the charged leptons
relative
to the beam such that $-0.95 \leq \cos \theta_{\ell \pm} \leq 0.95$
. This is the only cut we impose for the $\mu \tau$ and $\mu e$
final states.

One potential background is $\tau$ pair production with each of the $\tau$'s
decaying leptonically. At $\sqrt{s}$ of $200 \, GeV$, the four lepton processes
each have a cross section of around $1 \, pb$. This is to be compared to the
cross section for $\tau$ pair
production, about $5 \, pb$, multiplied by the branching ratios of $\tau$ into
$e$ or $\mu$ of $17.8\%$ each \cite{14}, yielding a rate into a final state
with
the same signature as we are considering of about $0.16 \, pb$. At higher
energies, the $\tau$ pair  production cross section is
falling like $1/s$ while the cross section for our processes remains large. In
addition, the $\tau$ pair process should have substantially greater missing
transverse momentum and energy with four neutrinos in the final state. It seems
 that this source of background is manageable.

The four lepton processes with one or more $\tau$'s in the final state ($\mu
\tau$ and $\tau \tau$) could feed down as a background to the processes without
any $\tau$ if the $\tau$(s) decays leptonically. However, factoring in the
$\tau$ decay branching ratio and accounting for the higher missing transverse
momentum and energy keeps this background under control.

Another potential background comes from two photon processes with the $e^{+}$
and $e^{-}$ undetected near the beam. This is relevant to the $\mu \mu$ and $e
e$ processes and we make a cut on missing transverse momentum to eliminate two
photon events as a background source; we require total visible ${p}_{T} > 10 \,
GeV$. We also require for these two processes that each charged lepton carry a
minimum energy, $E_{\ell} > 10 \, GeV$. Finally, again for the $\mu \mu$ and $e
e$ processes we make a cut on the invariant mass of the charged lepton pair; we
require $m_{\ell^{+} \ell -} > 25 \, GeV$ in order to eliminate the low
invariant mass dileptons corresponding to the photon pole in these processes.

In Fig. 4(a,b,c,d), we show the cross sections as a function of $\sqrt{s}$ for
the processes $e^{+}e^{-} \rightarrow \ell^{+} \ell^{\prime -} \nu
\bar{\nu}^{\prime}$ for $\ell^{+} \ell^{\prime -}$ equal to $\mu^{+} \tau^{-}$,
$\mu^{+} e^{-}$, $\mu^{+} \mu^{-}$, $e^{+} e^{-}$, respectively, with the cuts
as described above imposed. In each case, the solid line corresponds to the
case of standard model couplings, $\kappa_{\gamma} = \kappa_{Z} =1$, while the
dashed line is for $\kappa_{\gamma} = \kappa_{Z} = 0.9$, an example of a 10\%
deviation with the couplings set equal. The sensitivity to $\kappa_{V}$
increases with increasing center of mass energy.
 The $\mu
\tau$ process exhibits the most sensitivity to $\kappa_{V}$, as might be
expected
since it has the least number of extraneous contributing diagrams; however, it
also has the smallest cross section. Thus, it is useful to consider all the
processes.

We make our study of $\kappa_{V}$ dependence at two center of mass energies,
$500 \, GeV$ and $1 \, TeV$, motivated by the possibility of future high energy
$e^{+} e^{-}$ colliders. For each of the four types of four lepton processes,
at each of the two energies, we vary $\kappa_{\gamma}$ alone from 0.9 to 1.1,
$\kappa_{Z}$ alone over the same range, and $\kappa_{\gamma}$ constrained to
equal $\kappa_{Z}$ over the same range. As an example, we show the ratio of the
cross section with nonstandard couplings to the standard model cross section
for
the $\mu^{+} \tau^{-}$ process at $\sqrt{s}$ of $500 \, GeV$ and $1 \, TeV$ in
Figs. 5a and 5b, respectively; in each case, the solid line corresponds to
$\kappa_{\gamma}$ set equal to $\kappa_{Z}$, the dashed line to $\kappa_{Z} =
1$, and the dotted line to $\kappa_{\gamma} = 1$. For reference, the standard
model cross section in this case is 0.137 $pb$ at $\sqrt{s}$ of $500 \, GeV$
and
0.034 $pb$ at $\sqrt{s}$ of $1 \, TeV$. As in this example, we find that the
$\kappa_{\gamma} = \kappa_{Z}$ case always shows the greatest deviation from
the
standard model value of the cross section. At $\sqrt{s}$ of $500 \, GeV$, each
process is more sensitive to deviations of $\kappa_{V}$ below the standard
model
value of 1 than above it; however at the higher center of mass energy of $1 \,
TeV$, the sensitivity to $\kappa_{V}$ is considerably more symmetric about 1.
Varying either $\kappa_{\gamma}$ or $\kappa_{Z}$ separately or setting them
equal, the amplitude at each energy for each process can be expressed as $M =
\alpha + \beta \kappa$; we have in each case fit a parabola for the cross
section as a function of $\kappa$ and solved for the cross section as a
function
of the two parameters $\kappa_{\gamma}$ and $\kappa_{Z}$ as \begin{equation}
\sigma \sim |M|^{2} = a + b \kappa_{\gamma} + c \kappa_{Z} + d \kappa_{\gamma}
\kappa_{Z} + e \kappa_{\gamma}^{2} + f \kappa_{Z}^{2}. \end{equation} In Fig.
6(a,b,c,d), we show the resulting surface plots of the cross sections for the
$\mu \tau$, $\mu e$, $\mu \mu$, and $e e$ processes, respectively, as a
function
of $\kappa_{\gamma}$ and $\kappa_{Z}$ at $\sqrt{s}$ of $500 \, GeV$. The
corresponding results for $\sqrt{s}$ of $1 \, TeV$ are given in Fig. 7(a-d). We
have checked that a run over a grid of various ($\kappa_{\gamma}, \kappa_{Z}$)
values reproduces these results.

We turn these results into limits on the detection of deviations of
$\kappa_{V}$ from 1 by assuming an integrated luminosity of $50 \,fb^{-1}$ for
a proposed collider. \cite{15} Figs. 8(a,b,c,d) are $1\sigma$ and $2\sigma$
contour plots for the $\mu \tau$, $\mu e$, $\mu \mu$, and $e e$ processes,
respectively, at $500 \, GeV$ center of mass energy. The solid lines on each
plot are the $1\sigma$ contours and the dashed lines are the $2\sigma$
contours,
with statistical errors only included. The corresponding contours for
 $\sqrt{s} = 1 \, TeV$ are given in
Figs. 9(a-d). In obtaining these results, we have
included a factor of 2 to account for the charge conjugate processes in the
$\mu^{+} \tau^{-}$ ($\mu^{-} \tau^{+}$) and $\mu^{+} e^{-}$ ($\mu^{-} e^{+}$)
channels. The $\tau e$ and $\tau \tau$ channels would yield results as for the
$\mu e$ and $\mu \mu$ channels, respectively, the $\mu$ $\tau$ mass difference
being negligible here. Thus, from the total cross section of the individual
processes, we find the following $2\sigma$ limits on measurements of
$\kappa_{\gamma}$ and $\kappa_{Z}$. At $\sqrt{s}$ of $500 \, GeV$,
$\kappa_{\gamma}$ could be measured within $-2\% \,(\mu \tau)$ to $+7\% \,
(ee)$ and $\kappa_{Z}$ within the range $-4\% \, (\mu e)$ to $+7\% \, (\mu
e, \, ee)$. At $1 \, TeV$, the corresponding limits on $\kappa_{\gamma}$ are
$-0.7\% \, (\mu \tau)$ to $+2.8\% \, (\mu e)$ and on $\kappa_{Z}$ we find
limits
of $-1.5\% \, (\mu \tau)$ to $+2.3\% \, (\mu e)$. The channels given in
brackets
with each limit indicate which of the processes supplies the best bound. These
particular limits simply represent the outer bound of the $2\sigma$ contour
for the various processes. If one makes some assumptions about the
relationship of $\kappa_{\gamma}$ and $\kappa_{Z}$, such as that
$\kappa_{\gamma} = \kappa_{Z}$ or that $\Delta \kappa_{\gamma} = \frac{2
\cos^{2} \theta_{W}}{\cos^{2} \theta_{W} - \sin^{2} \theta_{W}} \Delta
\kappa_{Z}$ \cite{5}, better bounds (which can be read off Figs. 8 and 9) are
obtained. In addition, combining the statistics from all the processes
considered here would improve the bounds. In fact, one could also combine
these four lepton processes with the similar jet channels such as $e^{+} e^{-}
\rightarrow q \bar{q}^{\prime} \ell \nu$. Combined bounds would necessitate
inclusion of detector acceptances and efficiencies for the various particle
types. We emphasize that even the bounds quoted above from the cross sections
of individual processes are, indeed, approaching the very interesting realm of
probing $\kappa_{V}$ to within a few per cent of the standard model value. We
note that it is particularly important to go to the higher energy in order to
probe values of $\kappa_{V}$ larger than 1.

We have also generated a number of distributions; these include the
differential cross sections with respect to the angle of each charged lepton
relative to the beam, the angle between the charged leptons, the energy and
transverse momentum of each charged lepton, the total visible energy and
transverse momentum and the invariant mass of the charged lepton pair. The
angular distributions tend all to be quite strongly peaked along the beam line
for the standard model; they are generally enhanced somewhat away from the
beam direction for nonstandard $\kappa_{V}$ values. The energy
and transverse momentum distributions of the individual particles tend to be
enhanced over most of their range. The total visible transverse momentum is
preferentially enhanced where the differential cross section is largest. As
examples, we show in Figs. 10a and 10b the differential cross section with
respect to the total visible transverse momentum for the $\mu \tau$ and $\mu
\mu$ processes, respectively, both at a center of mass energy of $500 \, GeV$.
The solid line in each figure represents standard model couplings and the
dotted line is for the case of $\kappa_{\gamma} = \kappa_{Z} = 0.9$. The
differential cross section with respect to $x_{-}$, where $x_{-} = E_{-}/
(\sqrt{s}/2)$, is shown in Figs. 11a and 11b for the same two processes.
$E_{-}$
is the energy of the negatively charged final state lepton. The notation is
the same as for Fig. 10.

\section{Helicity Considerations}

Referring to the three-dimensional plots of Figs. (6a-d) and (7a-d), note
that a plane of constant cross section intersects a ring of $(\kappa_{\gamma},
\kappa_{Z})$ pairs. So we can apparently determine, within the limits given in
the last section, a deviation from the standard model with cross section
measurements but it remains to determine whether we can pinpoint the values of
$\kappa_{\gamma}$ and $\kappa_{Z}$ individually. There have been a number of
approaches proposed for discriminating between deviations of $\kappa_{\gamma}$
and $\kappa_{Z}$. One suggestion is to study processes which only involve one
or the other of the $\gamma W W$ and $Z W W$ vertices. The associated
production of a $W$ with either a $\gamma$ or a $Z$ boson, radiative
$W$ decay \cite{16}, and $e \gamma$ processes such as $ e \gamma \rightarrow W
\nu$ \cite{17} fall into this category. Another suggestion is to make cuts
which
isolate one of the vertices. For instance, Couture, Godfrey, and Lewis have
studied the $\mu^{+} \mu^{-}$ production process which we also consider here
and
have focussed on the $Z W W$ vertex by requiring that the invariant mass of the
$\mu^{+} \mu^{-}$ pair fall within $5 \, GeV$ of $M_{Z}$. \cite{18} This, in
effect, helps to isolate the $Z$ contribution from the diagram of the form
of Fig. 3.

Here, we emphasize instead the potential usefulness of the helicity structure
in
providing a determination of $\kappa_{\gamma}$ and $\kappa_{Z}$. For
instance, we can make one general statement regarding the contribution to
the total unpolarized cross section of the $(-++-)$ helicity. Recall that
this amplitude contributes to all our processes since it occurs for the $W$
pair type diagrams of Fig. (1), although it is not the dominant amplitude
(As previously noted, $(+-+-)$ is the dominant amplitude.). We observe that the
$(-++-)$ amplitude is suppressed at $\sqrt{s} \gg M_{Z}$ for $\kappa_{\gamma} =
\kappa_{Z}$ as a direct result of the general form of this amplitude, which
is given below.
\begin{equation}
M_{(-++-)} = \left[ \frac{\kappa_{\gamma}+1}{2s} - \frac{\kappa_{Z} + 1}
{2(s-M^{2}_{Z})} \right] \, A + B
\end{equation}
Here $A$ and $B$ denote the $\kappa_{V}$ dependent and independent
factors, respectively, of the amplitude. For large center of
mass energies, the cancellation of the $\kappa_{\gamma}$ and $\kappa_{Z}$
terms results in a $(-++-)$ helicity contribution of less than about
one per cent of the total cross section for the standard model and for
$\kappa_{\gamma} = \kappa_{Z}$ in general. On the other hand, for nonequal
values of $\kappa_{\gamma}$ and $\kappa_{Z}$, this contribution can be as
much as 30\% of the total. In Fig. 12, we illustrate this general
behaviour with examples for the $\mu \tau$ process. In this process, only
two helicity amplitudes contribute so presentation is simplified,
although the suppression of the $(-++-)$ is general for all the processes
as described above. We display the differential cross section with respect to
total visible transverse momentum for three sets of
$(\kappa_{\gamma},\kappa_{Z})$ values. Fig. 12(a,b,c) represent (1.0,1.0),
(0.9,0.9), and (0.9,1.0), respectively, at $\sqrt{s}$ of $1 \, TeV$. The solid
line corresponds to the unpolarized cross section; the dashed line corresponds
to the $(+-+-)$ helicity contribution and the dotted to the $(-++-)$
contribution. The $(-++-)$ amplitude is enhanced in Fig 12(c), where
$\kappa_{\gamma}$ is not equal to $\kappa_{Z}$. Thus, polarized beams accessing
the individual helicity contributions could differentiate between the
$\kappa_{\gamma} = \kappa_{Z}$ case and the nonequal case.

Apart from the general observation described above regarding the case of
$\kappa_{\gamma}$ and $\kappa_{Z}$ equal, experimental results on the cross
sections for the four types of processes we consider with polarized and
unpolarized beams could provide a characteristic `fingerprint' for a
$(\kappa_{\gamma},\kappa_{Z})$ pair. As an example of how this might work,
refer
to Fig. 5 for the $\mu \tau$ process at $1 \, TeV$; from that plot, we note
that, for instance, $(\kappa_{\gamma}, \kappa_{Z}) = (0.945,0.945), \,
(1.07,1.07), \, (1,1.095), \, (1,0.92)$, and $(0.92,1)$ all have approximately
the same total cross section. The percentage of the cross section supplied by
the $(-++-)$ helicity is  less than 1\% for the two cases quoted with
$\kappa_{\gamma}=\kappa_{Z}$; it is 3.6\% for $(1, 1.095)$, 18\% for $(1,
0.92)$, and 27\% for $(0.92,1)$. Since the total cross section for unpolarized
beams corresponds to about 4000 events, these cases can be discriminated
providing reasonable polarization can be achieved. In Fig. 13, we illustrate,
for the $\mu \tau$ process at $1 \, TeV$, the $(+-+-)$ and $(-++-)$ helicity
contributions to the differential cross section with respect to the normalized
$\tau$ energy, $x_{-}$, for $(\kappa_{\gamma},\kappa_{Z}) = (0.945,0.945)$
(solid
lines), $(1,0.92)$ (dashed lines), and $(0.92,1)$ (dotted lines). In
each case, the $(+-+-)$ helicity is the larger of the two corresponding
contributions and so is the upper line in each pair. For the
 $(\kappa_{\gamma},\kappa_{Z}) = (0.945,0.945)$ case, the $(-++-)$ contribution
is very small relative to the scale of the figure so it is marked also with
diamonds. The figure indicates the relative contribution from the different
helicities for the various values of $\kappa_{V}$. For simplicity of
presentation, we do not show the sum of the amplitudes but point out here that
not only are the total cross sections very similar for the various
$(\kappa_{\gamma}, \, \kappa_{Z})$ pairs but, in fact, the distributions for
unpolarized beams are as well; it is only for the various individual
polarization contributions that the $(\kappa_{\gamma}, \, \kappa_{Z})$ sets are
distinguished.  Similar results from the four types of processes can be
combined
to narrow in on the actual values of $\kappa_{\gamma}$ and $\kappa_{Z}$,
individually.

\section{Summary and Conclusions}

We have presented a study of the sensitivity to W boson coupling parameters,
$\kappa_{\gamma}$ and $\kappa_{Z}$, of the process $e^{+}e^{-}\rightarrow
\ell^{+} \ell^{\prime -} \nu
\bar{\nu}^{\prime}$, including the charged lepton final states $\mu
\tau$, $\mu e$ ($\tau e$), $\mu \mu$ ($\tau \tau$), and $ee$. The full matrix
element calculation has been performed for each of the four types of
processes. We find that, for a 500 $GeV$ $e^{+}e^{-}$ collider achieving an
integrated luminosity of 50 $fb^{-1}$, $\kappa_{\gamma}$ could be measured
within the limits from 0.98 to 1.07 at the $2 \sigma$ level and $\kappa_{Z}$
within the limits 0.96 to 1.07. For a 1 $TeV$ collider with the same
luminosity,
the corresponding limits are from 0.993 to 1.028 for $\kappa_{\gamma}$ and from
0.985 to 1.023 for $\kappa_{Z}$. These limits are all for total cross section
measurements of individual reactions. The 1 $TeV$ limits, in particular, are
very interesting even at the level of standard model radiative corrections;
the higher energy is particularly important in determining $\kappa_{V}$
values which may be greater than the standard model tree level value of 1.

We have also found that beam polarization would be useful in determining
values of $\kappa_{\gamma}$ and $\kappa_{Z}$ individually as opposed to
merely a deviation of either parameter from the standard model value. For all
the processes, the helicity amplitude $(-++-)$ is suppressed in the case that
$\kappa_{\gamma}$ and $\kappa_{Z}$ are equal at the high energies considered
here. Thus, for instance, if the dominant $(+-+-)$ helicity contributed
within about a per cent of the total cross section for the $\mu \tau$
process, equality of $\kappa_{\gamma}$ and $\kappa_{Z}$ would be indicated.
On the other hand, for nonequal values, the $(+-+-)$ contribution might be as
little as 70\%. Thus, polarized beams could determine the contributions of
the various helicity amplitudes and yield values of
$\kappa_{\gamma}$ and $\kappa_{Z}$ individually.

In conclusion, the processes considered here offer a very clean experimental
signature for excellent sensitivity to $\kappa_{\gamma}$ and $\kappa_{Z}$ at
a high energy $e^{+}e^{-}$ collider.

\newpage
\begin{center}
ACKNOWLEDGEMENTS
\end{center}

This work was funded in part by the Natural Sciences and Engineering Council of
Canada. P.K. gratefully acknowledges the hospitality of the Phenomenology
Institute at the University of Wisconsin at Madison and useful discussions
with Dieter Zeppenfeld. The authors also thank Stephen Godfrey for helpful
discussions.

\bibliographystyle{unsrt}

\begin{thebibliography}{99}
\bibitem{1} M. Timko, `Status of $W \gamma$ Search at CDF', contribution to
{\em
 The Spring Meeting of the American Physics Society}, Washington, D.C.,
16-19 April 1990 (unpublished).
\bibitem{2} J. Alitti {\em et al.}, Phys. Lett. {\bf 277B}, 194 (1992).
\bibitem{3} A. de Rujula {\em et al.}, CERN TH.6272 (1991; unpublished); D.
Choudhury, P. Roy, and R. Sinha, TIFR-TH/93-08.
\bibitem{4} C. Burgess and D.
London, McGill University reports McGill-92-04; McGill-92-05; McGill-92-14
(1992;
unpublished).
\bibitem{5} K. Hagiwara {\em et al.}, Phys. Lett. {\bf283B}, 353
(1992).
\bibitem{6} K. Hagiwara {\em et al.}, Nucl. Phys. {\bf B282}, 253
(1987).
\bibitem{7} J.F.Gaemers and G.J. Gounaris, Z. Phys. {\bf C1}, 259
(1979).
\bibitem{8} W. Marciano and A. Queijeiro, Phys. Rev. D
{\bf33}, 3449 (1986); F. Boudjema {\em et al.}, Phys. Rev. D {\bf43}, 2223
(1991).
\bibitem{9} T.G. Rizzo and M.A. Samuel, Phys. Rev. D {\bf35}, 403 (1987).
\bibitem{10} G. Couture {\em et al.}, Phys. Rev. D {\bf 36}, 859 (1987).
\bibitem{11} G. Couture {\em et al.}, Phys. Rev. D {\bf38}, 860 (1988).
\bibitem{12} R. Kleiss and W.J. Stirling, Nucl. Phys. {\bf B262}, 235 (1985).
\bibitem{13} F. Bletzacker and H.T. Nieh, Nucl. Phys. {\bf B124}, 511 (1977);
P.
Mery and M. Perrottet, Nucl. Phys. {\bf B175}, 234 (1980); C.L. Bilchak and
J.D. Stroughair, Phys. Rev. D {\bf30}, 1881 (1984).
\bibitem{14} M. Aguilar-Benitez {\em et al.}, Phys. Rev. D {\bf45}, Part 2
(1992).
\bibitem{15} This estimated integrated luminosity is based on information in
{\em Proceedings of the Third Workshop on the Japan Linear
Collider}, KEK, Japan, 1992, edited by A. Miyamoto; {\em Search for New
Phenomena at Colliding Beam Facilities}, New Haven, Connecticut, 1992, to be
published.
\bibitem{16} K.O. Mikaelian, M.A. Samuel, and D. Sahdev, Phys. Rev.
Lett. {\bf43}, 746 (1979); J. Cortes, K. Hagiwara, and F. Herzog, Nucl. Phys.
{\bf B278}, 16 (1986); U. Baur and D. Zeppenfeld, Nucl. Phys. {\bf B308}, 127
(1988); U. Baur and E. Berger, Phys. Rev. D {\bf41}, 1476 (1990); M. Samuel
{\em
et al.}, Phys. Lett. {\bf280B}, 124 (1992).  \bibitem{17} G. Couture, S.
Godfrey,
and P. Kalyniak, Phys. Lett. {\bf B218}, 361 (1989); Phys.Rev. D {\bf39}, 3239
(1989); D {\bf42}, 1841 (1990).   \bibitem{18} G. Couture, S. Godfrey, and R.
Lewis, Phys. Rev. D {\bf45}, 777 (1992).   \end{thebibliography}

\begin{figure}
\caption{The $W$ pair production diagrams which are relevant to all the
processes considered.}
\end{figure}

\begin{figure}
\caption{The diagrams with one $W$ and either a $\gamma$ or $Z$ in the
t-channel. Fig. 2a contributes to the $\mu^{+} e^{-}$ and $e^{+} e^{-}$
processes and Fig. 2b to the $e^{+} e^{-}$ process.}
\end{figure}

\begin{figure}
\caption{The diagram with two $W$ bosons in the t-channel contributes to the
$\mu^{+} \mu^{-}$ and $e^{+} e^{-}$ processes.}
\end{figure}

\begin{figure}
\caption{The total cross section as a function of center of mass energy for the
a) $\mu^{+} \tau^{-}$, b) $\mu^{+} e^{-}$, c) $\mu^{+} \mu^{-}$, and d) $e^{+}
e^{-}$ processes. The solid line in each figure corresponds to the standard
model case while the dashed lines are for $\kappa_{\gamma}= \kappa_{Z}=0.9$
.}
\end{figure}

\begin{figure}
\caption{For the $\mu \tau$ process at a) $500 \, GeV$ and b) $1 \, TeV$, the
ratio of the cross section for nonstandard values of $\kappa_{V}$ to the
standard model cross section as a function of $\kappa_{V}$. The solid line
corresponds to $\kappa_{\gamma} = \kappa_{Z}$, the dashed line to $\kappa_{Z} =
1$, and the dotted line to $\kappa_{\gamma} = 1$ in parts a and b.}
\end{figure}

\begin{figure}
\caption{At the center of mass energy of $500 \, GeV$, the cross section as a
function of $\kappa_{\gamma}$ and $\kappa_{Z}$ for the a) $\mu \tau$, b) $\mu
e$, c) $\mu \mu$, and d) $e e$ processes.}
\end{figure}

\begin{figure}
\caption{At the center of mass energy of $1 \, TeV$, the cross section as a
function of $\kappa_{\gamma}$ and $\kappa_{Z}$ for the a) $\mu \tau$, b) $\mu
e$, c) $\mu \mu$, and d) $e e$ processes.}
\end{figure}

\begin{figure}
\caption{Contour plots in $\kappa_{\gamma}$ and $\kappa_{Z}$ at the $1\sigma$
(solid lines) and $2\sigma$ (dashed lines) levels for the a) $\mu \tau$, b)
$\mu
e$, c) $\mu \mu$, and d) $e e$ processes at $500 \, GeV$.}
\end{figure}

\begin{figure}
\caption{Contour plots in $\kappa_{\gamma}$ and $\kappa_{Z}$ at the $1\sigma$
(solid lines) and $2\sigma$ (dashed lines) levels for the a) $\mu \tau$, b)
$\mu
e$, c) $\mu \mu$, and d) $e e$ processes at $1 \, TeV$.}
\end{figure}

\begin{figure}
\caption{The differential cross section with respect to total visible
transverse momentum at a center of mass energy of $500 \, GeV$ for the a) $\mu
\tau$ and b) $\mu \mu$ processes. The solid line corresponds to the standard
model while the dotted line is for $\kappa_{\gamma} = \kappa_{Z} = 0.9$.}
\end{figure}

\begin{figure}
\caption{The differential cross section with respect to the normalized energy
variable of the negative lepton, $x_{-}$, at a center of mass energy of $500 \,
GeV$ for the a) $\mu \tau$ and b) $\mu \mu$ processes. The solid line
corresponds to the standard model while the dotted line is for $\kappa_{\gamma}
= \kappa_{Z} = 0.9$.}
\end{figure}

\begin{figure}
\caption{For the $\mu \tau$ process at the center of mass energy of $1 \, TeV$,
the differential cross section with respect to the total visible transverse
momentum for a) $\kappa_{\gamma} = \kappa_{Z} =1$, b) $\kappa_{\gamma} =
\kappa_{Z} = 0.9$, and c) $\kappa_{\gamma} = 0.9$ with $\kappa_{Z} = 1$. The
solid line is the sum of all helicity amplitude contributions; the dashed line
is the $(+-+-)$ contribution and the dotted line is the $(-++-)$ contribution.}
\end{figure}

\begin{figure}
\caption{For the $\mu \tau$ process at $1 \, TeV$, the $(+-+-)$ and $(-++-)$
contributions to the differential cross section with respect to the normalized
$\tau$ energy, $x_{-}$, for $(\kappa_{\gamma},\kappa_{Z}) = (0.945,0.945)$
(solid
lines), (1,0.92) (dashed lines), and (0.92,1) (dotted lines). In each pair of
lines, the upper line represents the dominant $(+-+-)$ contribution; for the
(0.945,0.945) case, the $(-++-)$ contribution is very small on the scale of
the figure and is marked with diamonds.}
\end{figure}

\end{document}